\documentclass[aps,preprint]{revtex4}%
\usepackage{amsfonts}
\usepackage{amsmath}
\usepackage{amssymb}
\usepackage{graphicx}%
\newcommand {\be}{\begin{equation}}
\newcommand {\ee}{\end{equation}}

\begin{document}
\title{Neutron matter on the lattice with pionless effective field theory}
\author{Dean Lee}
\affiliation{Department of Physics, North Carolina State University, Raleigh, NC 27695}
\author{Thomas Sch{\"a}fer}
\affiliation{Department of Physics, North Carolina State
University,\ Raleigh,\ NC\ 27695\linebreak RIKEN-BNL Research Center,
Brookhaven National Laboratory, Upton, NY 11973}
\keywords{nuclear lattice simulation non-perturbative chiral effective field theory}
\pacs{21.30-x,21.65+f,13.75.Cs}

\begin{abstract}
We study neutron matter by combining pionless effective field theory with
non-perturbative lattice methods. \ The neutron contact interaction is
determined by zero temperature scattering data. \ We simulate neutron matter
on the lattice at temperatures 4 and 8 MeV and densities below one-fifth
normal nuclear matter density. \ Our results at different lattice spacings
agree with one another and match bubble chain calculations at low densities.
\ The equation of state of pure neutron matter obtained from our simulations
agrees quantitatively with variational calculations based on realistic potentials.

\end{abstract}
\maketitle

\section{Introduction}

The equation of state of dilute neutron matter is of central importance to the
structure and evolution of neutron stars
\cite{Heiselberg:2000dn,Lattimer:2000nx}. In addition to that, the neutron
matter problem has many interesting physical aspects. Neutron matter has
positive pressure at all densities and becomes a superfluid at sufficiently
small temperature. If the density is small, pairing is expected to take place
in an $S$-wave, but at higher density $P$-wave pairing might be dominant
\cite{Dean:2002zx}. The neutron matter problem contains a number of very
different scales. The neutron scattering length is very large, $a_{nn}%
\simeq-18$ fm, which implies that the dimensionless parameter $k_{F}%
|a_{nn}|\gg1$ for densities $\rho>10^{-4}\rho_{N}$. Here, $k_{F}=(3\pi^{2}%
\rho)^{1/3}$ is the Fermi momentum and $\rho_{N}\simeq0.17$ fm$^{-3}$ is the
saturation density of nuclear matter. The effective range, on the other hand,
is of natural size, $r_{nn}\simeq2.8$ fm. As a consequence, the parameter
$k_{F}|r_{nn}|$ is neither large nor small for densities $\rho\sim\rho_{N}$.

If the density is very small, $\rho<0.1\rho_{N}$, then $k_{F}|r_{nn}|$ is a
small parameter and neutron matter is close to the limit in which
$k_{F}|a_{nn}|\rightarrow\infty$ and $k_{F}|r_{nn}|\rightarrow0$. In this
limit dimensional analysis implies that the energy per particle and the gap
have to be proportional to the Fermi energy
\begin{equation}
\frac{E}{A}=\xi\frac{3}{5}\frac{k_{F}^{2}}{2m},\hspace{0.4cm}\Delta=\zeta
\frac{k_{F}^{2}}{2m}. \label{Bertsch}%
\end{equation}
The determination of the two dimensionless parameters $\xi$ and $\zeta$ is a
fascinating non-perturbative problem that has received a lot of attention
recently. This interest was fueled by experimental advances in creating cold,
dilute gases of fermionic atoms tuned to be near a Feshbach resonance
\cite{O'Hara:2002,Gupta:2002,Regal:2003,Bourdel:2003,Gehm:2003}.

The traditional approach to the neutron matter problem is based on the
assumption that nucleons can be treated as non-relativistic point-particles
interacting mainly via two-body potentials. The two-body potentials are fitted
to experimental data on nucleon-nucleon scattering. The many-body problem is
addressed by solving the many-body Schr\"{o}dinger equation using variational
methods or Green function Monte Carlo methods guided by variational wave
functions \cite{Friedman:1981qw,Carlson:2003wm}.

Even though this method has been very successful it is desirable to seek an
alternative approach that is more directly related to QCD, systematically
improvable, and that lends itself to numerical studies which do not rely on
variational wave functions. Such an approach is provided by effective field
theory. The use of effective field theory (EFT) methods in nuclear physics was
pioneered by Weinberg \cite{Weinberg:1990rz}. Over the last few years EFT
methods have been applied successfully to the study of two and three-body
systems at low energy \cite{Epelbaum:1998na,Beane:2000fx,Bedaque:2002mn}.
Nuclear and neutron matter was studied using a perturbative expansion in
powers of the Fermi momentum \cite{Kaiser:2001jx} and using lattice
simulations \cite{Muller:1999cp,Lee:2004si}.

In this paper we shall study dilute neutron matter using a nuclear effective
field theory on the lattice. Since we are interested in densities below
nuclear matter saturation density we shall assume that the relevant momenta
are smaller than the pion mass and that we can use an effective field theory
that contains only neutrons. In this work we shall limit ourselves to the
lowest order effective Lagrangian which contains a single four-fermion contact
interaction with no derivatives that is adjusted to the neutron-neutron
scattering length. This effective theory is sufficient in order to investigate
universal properties in the limit $k_{F}|a_{nn}|\rightarrow\infty$,
$k_{F}|r_{nn}|\rightarrow0$. An important advantage of the model is the fact
that in the case of an attractive interaction there is no sign problem at
finite density \cite{Chen:2003vy}. As a consequence, the theory can be
simulated efficiently using standard hybrid Monte Carlo algorithms
\cite{Wingate:2004wm}.

The paper is organized as follows. In Sects. II-IV we introduce the lattice
theory. In Sect.~V we discuss how to determine the coefficient of the four
fermion interaction by matching to the two-body scattering length. In Sect.~VI
we study a low density approximation to the partition function based on
summing particle-particle chains. In Sect.~VII we describe our hybrid Monte
Carlo method. Numerical results for the neutron density, the energy per
particle and the equation of state are given in Sects. VIII-XII.

\section{Notation}

Before describing the physics we first define some notation we use throughout
our discussion. \ We let $\vec{n}$ represent integer-valued lattice vectors on
our $3+1$ dimensional space-time lattice. \ We use a subscripted
\textquotedblleft$s$\textquotedblright\ such as in $\vec{n}_{s}$ to represent
purely spatial lattice vectors. \ We use subscripted indices such as $i,j$ for
the two spin components of the neutron, $\uparrow$ and $\downarrow$. \ We let
$\hat{0}$ be the unit lattice vector in the time direction and let $\hat
{l}_{s}=\hat{1}$, $\hat{2}$, $\hat{3}$ be the corresponding unit lattice
vectors in the spatial directions. \ A summation symbol such as
\begin{equation}
\sum_{l_{s}}%
\end{equation}
implies a summation over values $l_{s}=1$, $2$, $3$.

We take the neutron mass to be $939$ MeV, and normal nuclear matter density to
be $0.17$ fm$^{-3}$. \ We let $a$ be the lattice spacing in the spatial
direction and $L$ be the number of lattice sites in each spatial direction.
\ $a_{t}$ is the lattice spacing in the temporal direction and $L_{t}$ is the
number of lattice sites in the temporal direction. \ We let $\alpha_{t}$ be
the ratio between lattice spacings,%
\begin{equation}
\alpha_{t}=\tfrac{a_{t}}{a}.
\end{equation}
Throughout we use dimensionless parameters and operators, which correspond
with physical values multiplied by the appropriate power of $a$. \ In the end,
however, we report final results in physical units such as MeV or fm$^{-3}$.
\ In cases where there may be confusion, we use the subscript $phys$ to
identify quantities in physical units.

We use $a,a^{\dagger}$ to represent annihilation and creation operators for
the neutron, whereas $c,c^{\ast}$ indicate the corresponding Grassmann
variables in the path integral representation. \ We let $m_{N}$ be the mass of
the neutron and $\mu$ be the neutron chemical potential. \ For the neutron
fields we apply periodic boundary conditions in the spatial directions and
antiperiodic boundary conditions in the temporal direction. \ For each neutron
momentum we use the notation%
\begin{equation}
\vec{k}_{\ast}=(\tfrac{2\pi}{L_{t}}k_{0},\tfrac{2\pi}{L}k_{1},\tfrac{2\pi}%
{L}k_{2},\tfrac{2\pi}{L}k_{3}),
\end{equation}
where $k_{1}$, $k_{2}$, and $k_{3}$ are integers and $k_{0}$ is an odd
half-integer. \ In physical units the momentum is
\begin{equation}
\vec{k}_{phys}=(k_{\ast0}a_{t}^{-1},k_{\ast1}a^{-1},k_{\ast2}a^{-1},k_{\ast
3}a^{-1}).
\end{equation}
Unless otherwise indicated, our the momentum labels will follow this
convention. \ For convenience we also define%
\begin{equation}
h=\tfrac{\alpha_{t}}{2m_{N}},
\end{equation}
and%

\begin{equation}
\omega_{k}=6h-2h\sum_{l_{s}}\cos(k_{\ast l_{s}}). \label{omega}%
\end{equation}

We let $D^{free}(\vec{k})\delta_{ij}$ be the free neutron propagator. \ For
notational convenience the spin-conserving $\delta_{ij}$ in the neutron
propagator will be implicit. \ The self-energy, $\Sigma(\vec{k})$, is defined
by%
\begin{equation}
D^{full}(\vec{k})=\frac{D^{free}(\vec{k})}{1-\Sigma(\vec{k})D^{free}(\vec{k}%
)},
\end{equation}
where $D^{full}(\vec{k})$ is\ the fully-interacting propagator.

In our plots we use the abbreviation \textquotedblleft fc\textquotedblright%
\ for free continuum results, \textquotedblleft f\textquotedblright\ for free
lattice results, \textquotedblleft b\textquotedblright\ for bubble chain
calculations, and \textquotedblleft s\textquotedblright\ for lattice
simulations results. \ In addition to these abbreviations, we will use the
shorthand labels shown in Table 1 for various combinations of spatial and
temporal lattice spacings presented in our analysis.%
\[%
\genfrac{}{}{0pt}{0}{\text{Table 1: Shorthand labels for various lattice
spacings used}}{%
\begin{tabular}
[c]{|l|l|l|}\hline
$a^{-1}($MeV$)$ & $a_{t}^{-1}($MeV$)$ & Label\\\hline
$50$ & $24$ & $0$\\\hline
$60$ & $32$ & $1$\\\hline
$60$ & $48$ & $2$\\\hline
$70$ & $64$ & $3$\\\hline
$80$ & $72$ & $4$\\\hline
\end{tabular}
}%
\]

\section{Free nucleon}

On the lattice the free neutron Hamiltonian can be written as%
\begin{align}
H_{\bar{N}N}  &  =\sum_{\vec{n}_{s},i}\left[  (m_{N}-\mu+\tfrac{3}{m_{N}%
})a_{i}^{\dagger}(\vec{n}_{s})a_{i}(\vec{n}_{s})\right] \nonumber\\
&  -\tfrac{1}{2m_{N}}\sum_{\vec{n}_{s},l_{s},i}\left[  a_{i}^{\dagger}(\vec
{n}_{s})a_{i}(\vec{n}_{s}+\hat{l}_{s})+a_{i}^{\dagger}(\vec{n}_{s})a_{i}%
(\vec{n}_{s}-\hat{l}_{s})\right]  .
\end{align}
We can approximate the partition function as a Euclidean lattice path
integral,%
\begin{equation}
Z_{G}^{free}=Tr\exp\left[  -\beta H_{\bar{N}N}\right]  \simeq z_{0}^{free}\int
DcDc^{\ast}\exp\left[  -S^{free}\right]  , \label{basicpath}%
\end{equation}
where $z_{0}^{free}$ is a constant and%
\begin{align}
S^{free}  &  =\sum_{\vec{n},i}\left[  c_{i}^{\ast}(\vec{n})c_{i}(\vec{n}%
+\hat{0})-e^{-(m_{N}-\mu)\alpha_{t}}(1-6h)c_{i}^{\ast}(\vec{n})c_{i}(\vec
{n})\right] \nonumber\\
&  -he^{-(m_{N}-\mu)\alpha_{t}}\sum_{\vec{n},l_{s},i}\left[  c_{i}^{\ast}%
(\vec{n})c_{i}(\vec{n}+\hat{l}_{s})+c_{i}^{\ast}(\vec{n})c_{i}(\vec{n}-\hat
{l}_{s})\right]  .
\end{align}
We have taken a slightly different form than that used in \cite{Lee:2004si}.
\ Instead of the$\ e^{-6h}$ that appears in \cite{Lee:2004si}, we use the more
standard $1-6h$ as the coefficient multiplying $c_{i}^{\ast}(\vec{n}%
)c_{i}(\vec{n})$.

It is conventional to define a new normalization for $c_{i}$,
\begin{equation}
c_{i}^{\prime}=c_{i}e^{-(m_{N}-\mu)\alpha_{t}}.
\end{equation}
Then
\begin{equation}
Z_{G}^{free}\simeq z_{0}^{free}e^{-2(m_{N}-\mu)\beta L^{3}}\int Dc^{\prime
}Dc^{\ast}\exp\left[  -S^{free}\right]
\end{equation}
where%
\begin{align}
S^{free}  &  =\sum_{\vec{n},i}\left[  e^{(m_{N}-\mu)\alpha_{t}}c_{i}^{\ast
}(\vec{n})c_{i}^{\prime}(\vec{n}+\hat{0})-(1-6h)c_{i}^{\ast}(\vec{n}%
)c_{i}^{\prime}(\vec{n})\right] \nonumber\\
&  -h\sum_{\vec{n},l_{s},i}\left[  c_{i}^{\ast}(\vec{n})c_{i}^{\prime}(\vec
{n}+\hat{l}_{s})+c_{i}^{\ast}(\vec{n})c_{i}^{\prime}(\vec{n}-\hat{l}%
_{s})\right]  . \label{free}%
\end{align}

In momentum space we have%
\begin{equation}
S^{free}=\sum_{\vec{k},i}\tilde{c}_{i}^{\ast}(-\vec{k})\tilde{c}_{i}^{\prime
}(\vec{k})\left[  e^{-ik_{\ast0}+(m_{N}-\mu)\alpha_{t}}-(1-6h)-2h\sum_{l_{s}%
}\cos(k_{\ast l_{s}})\right]  .
\end{equation}
The free neutron correlation function on the lattice is%
\begin{equation}
\frac{\int Dc^{\prime}Dc^{\ast}c_{i}^{\prime}(\vec{n})c_{i}^{\ast}%
(0)\exp\left[  -S^{free}\right]  }{\int Dc^{\prime}Dc^{\ast}\exp\left[
-S^{free}\right]  }=\frac{1}{L_{t}L^{3}}\sum_{\vec{k}}e^{-i\vec{k}_{\ast}%
\cdot\vec{n}}D^{free}(\vec{k}), \label{pathcorrelator}%
\end{equation}
(no sum over $i$) where the free neutron propagator is%
\begin{align}
D^{free}(\vec{k})  &  =\frac{1}{e^{-ik_{\ast0}+(m_{N}-\mu)\alpha_{t}%
}-(1-6h)-2h\sum_{l_{s}}\cos(k_{\ast l_{s}})}\nonumber\\
&  =\frac{1}{e^{-ik_{\ast0}+(m_{N}-\mu)\alpha_{t}}-1+\omega_{k}}%
\end{align}

\section{Neutron contact term}

There are two contact interactions at lowest order in the effective theory of
nucleons without pions. \ But since we are considering pure neutron matter,
this reduces to one contact interaction of the form%
\begin{equation}
H_{\bar{N}N\bar{N}N}=C\sum_{\vec{n}_{s}}a_{\uparrow}^{\dagger}(\vec{n}%
_{s})a_{\uparrow}(\vec{n}_{s})a_{\downarrow}^{\dagger}(\vec{n}_{s}%
)a_{\downarrow}(\vec{n}_{s}).
\end{equation}
Since
\begin{equation}
\exp\left[  -\frac{C\alpha_{t}}{2}(a_{\uparrow}^{\dagger}a_{\uparrow
}+a_{\downarrow}^{\dagger}a_{\downarrow})^{2}\right]  =\sqrt{\frac{1}{2\pi}%
}\int_{-\infty}^{\infty}ds\exp\left[  -\frac{1}{2}s^{2}+s\sqrt{-C\alpha
}(a_{\uparrow}^{\dagger}a_{\uparrow}+a_{\downarrow}^{\dagger}a_{\downarrow
})\right]  ,
\end{equation}
we can write%
\begin{equation}
\exp\left[  -C\alpha_{t}a_{\uparrow}^{\dagger}a_{\uparrow}a_{\downarrow
}^{\dagger}a_{\downarrow}\right]  =\sqrt{\frac{1}{2\pi}}\int_{-\infty}%
^{\infty}ds\exp\left[  -\frac{1}{2}s^{2}+\left(  s\sqrt{-C\alpha}%
+\frac{C\alpha_{t}}{2}\right)  (a_{\uparrow}^{\dagger}a_{\uparrow
}+a_{\downarrow}^{\dagger}a_{\downarrow})\right]  .
\end{equation}
With this interaction the partition function can be approximated by
\begin{equation}
Z_{G}=Tr\exp\left[  -\beta(H_{\bar{N}N}+H_{\bar{N}N\bar{N}N})\right]  \simeq
z_{0}\int DsDcDc^{\ast}\exp\left[  -S\right]  ,
\end{equation}
where $z_{0}$ is a constant and%

\begin{align}
S  &  =\sum_{\vec{n},i}\left[  e^{(m_{N}-\mu)\alpha_{t}}c_{i}^{\ast}(\vec
{n})c_{i}^{\prime}(\vec{n}+\hat{0})-e^{\sqrt{-C\alpha_{t}}s(\vec{n}%
)+\frac{C\alpha_{t}}{2}}(1-6h)c_{i}^{\ast}(\vec{n})c_{i}^{\prime}(\vec
{n})\right] \nonumber\\
&  -h\sum_{\vec{n},l_{s},i}\left[  c_{i}^{\ast}(\vec{n})c_{i}^{\prime}(\vec
{n}+\hat{l}_{s})+c_{i}^{\ast}(\vec{n})c_{i}^{\prime}(\vec{n}-\hat{l}%
_{s})\right]  +\frac{1}{2}\sum_{\vec{n}}s^{2}(\vec{n}).
\end{align}

This lattice action is quite simple, and in the future it may be worth
considering improved actions in order to reduce discretization errors.
\ Nevertheless our lattice action maintains some important properties. \ One
property is that the chemical potential, $\mu$, is coupled to an exactly
conserved neutron number operator. \ This is clear since $\mu$ appears in the
same manner as a temporal gauge link. \ Another feature is that in the limit
as $m_{N}\rightarrow\infty$, we find%
\begin{equation}
Tr\exp\left[  -\beta(H_{\bar{N}N}+H_{\bar{N}N\bar{N}N})\right]  =z_{0}\int
DsDcDc^{\ast}\exp\left[  -S\right]  +O(m_{N}^{-2})\text{.}%
\end{equation}
Therefore any dependence on the temporal lattice spacing is suppressed by a
factor of $m_{N}^{-2}$. \ This makes it possible to take the static neutron
limit as a precision test of the simulation results. \ We have found this test
quite useful in the process of code development and checking.

\section{Determining coefficients}

\begin{figure}
[ptb]
\begin{center}
\includegraphics[
height=0.8216in,
width=2.1136in
]%
{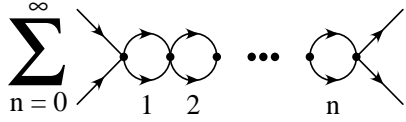}%
\caption{Bubble chain diagrams contributing to neutron-neutron scattering.}%
\label{scatt}%
\end{center}
\end{figure}

The interaction coefficient $C$ must be determined for various lattice
spacings $a$ and $a_{t}$. \ We do this by summing all bubble chain diagrams
contributing to neutron-neutron scattering as shown in Fig. \ref{scatt}.%

The next step is to locate the pole in the scattering amplitude and compare
with L\"{u}scher's formula for energy levels in a finite periodic box
\cite{Luscher:1986pf,Beane:2003da},%

\begin{equation}
E_{0}=\dfrac{4\pi a_{scatt}}{m_{N}L^{3}}[1-c_{1}\frac{a_{scatt}}{L}+c_{2}%
\frac{a_{scatt}^{2}}{L^{2}}+\cdots], \label{lus}%
\end{equation}
where $c_{1}=-2.837297,$ $c_{2}=6.375183$.\ \ We then tune the coefficient $C$
to give the physically measured $^{1}S_{0}$ scattering length. \ Since this
scattering length is much larger than any other length scale, we are in
essence probing the universal behavior of interacting fermions at infinite
scattering length. \ For our results we have used $a_{scatt}^{^{1}S_{0}}%
\simeq-24$ fm, though using the value of $-18$ fm specific for neutron-neutron
scattering changes the operator coefficient by only $1\%$.

The full bubble chain will have a pole when the amplitude for a single bubble
times one vertex coefficient equals $1$. \ We take the center of mass frame
and let the total incoming momentum of the two neutrons in physical units be
\begin{equation}
\vec{p}_{phys}=(p_{\ast0}a_{t}^{-1},0,0,0).
\end{equation}
Since the physical pole occurs in Minkowski space, in the end $p_{0}$ will be
imaginary. \ If we set $\mu=0$, then the amplitude for one bubble times one
vertex coefficient is%

\begin{equation}
(1-6h)^{2}\left(  e^{-C\alpha_{t}}-1\right)  B(p_{0}), \label{amplitude}%
\end{equation}
where%
\begin{equation}
B(p_{0})=\frac{1}{L^{3}L_{t}}\sum_{\vec{k}}\frac{1}{e^{m_{N}\alpha_{t}%
}e^{-ip_{\ast0}/2}e^{-ik_{\ast0}}-1+\omega_{k}}\frac{1}{e^{m_{N}\alpha_{t}%
}e^{-ip_{\ast0}/2}e^{ik_{\ast0}}-1+\omega_{k}}, \label{B}%
\end{equation}
and the condition for the location of the pole is%
\begin{equation}
B(p_{0})=\frac{1}{(1-6h)^{2}\left(  e^{-C\alpha_{t}}-1\right)  }.
\end{equation}

By the definition of $\omega_{k}$ in (\ref{omega})\ we see that $0\leq
\omega_{k}\leq12h$. \ We assume that the lattice spacing in the temporal
direction is sufficiently small so that $h\leq\frac{1}{6}$. \ In practice this
presents no problem since $m_{N}$ is quite large. \ We then have
\begin{align}
0  &  \leq\omega_{k}\leq2,\\
-1  &  \leq1-\omega_{k}\leq1.
\end{align}
We now make a variable transformation,%
\begin{equation}
z=e^{-ik_{\ast0}}=e^{-i\frac{2\pi}{L_{t}}k_{0}}.
\end{equation}
We also take the zero temperature limit, $L_{t}\rightarrow\infty$, and convert
from the discrete sum over $k_{0}$ to an integral clockwise over the unit
circle in $z$ using%
\begin{equation}
dz=-i\frac{2\pi}{L_{t}}z\,dk_{0}\text{,}%
\end{equation}%
\begin{equation}
dk_{0}=i\frac{L_{t}}{2\pi z}dz\text{.}%
\end{equation}
We then find%
\begin{align}
B(p_{0})  &  =\frac{i}{2\pi L^{3}}\sum_{k_{1},k_{2},k_{3}}\oint\frac
{dz}{z\left(  e^{m_{N}\alpha_{t}}e^{-ip_{\ast0}/2}z-1+\omega_{k}\right)
\left(  e^{m_{N}\alpha_{t}}e^{-ip_{\ast0}/2}z^{-1}-1+\omega_{k}\right)
}\nonumber\\
&  =\frac{i}{2\pi L^{3}}\sum_{k_{1},k_{2},k_{3}}\oint\frac{dz}{\left(
e^{m_{N}\alpha_{t}}e^{-ip_{\ast0}/2}z-(1-\omega_{k})\right)  \left(
e^{m_{N}\alpha_{t}}e^{-ip_{\ast0}/2}-\left(  1-\omega_{k}\right)  z\right)
}\nonumber\\
&  =-\frac{i}{2\pi L^{3}}\sum_{k_{1},k_{2},k_{3}}\oint\frac{e^{-m_{N}%
\alpha_{t}}e^{ip_{\ast0}/2}\left(  1-\omega_{k}\right)  ^{-1}dz}{\left(
z-e^{-m_{N}\alpha_{t}}e^{ip_{\ast0}/2}(1-\omega_{k})\right)  \left(
z-e^{m_{N}\alpha_{t}}e^{-ip_{\ast0}/2}\left(  1-\omega_{k}\right)
^{-1}\right)  }.
\end{align}
When $\operatorname{Re}(ip_{\ast0}\alpha_{t}^{-1})$ $<2m_{N}$ we pick up the
residue at $e^{-m_{N}\alpha_{t}}e^{ip_{\ast0}/2}(1-\omega_{k})$, and the
amplitude is
\begin{align}
B(p_{0})  &  =-\frac{i}{2\pi L^{3}}\sum_{k_{1},k_{2},k_{3}}\frac{-2\pi
i\left(  1-\omega_{k}\right)  ^{-1}}{(1-\omega_{k})-e^{2m_{N}\alpha_{t}%
}e^{-ip_{\ast0}}\left(  1-\omega_{k}\right)  ^{-1}}\nonumber\\
&  =-\frac{i}{2\pi L^{3}}\sum_{k_{1},k_{2},k_{3}}\frac{-2\pi i}{(1-\omega
_{k})^{2}-e^{2m_{N}\alpha_{t}}e^{-ip_{\ast0}}}\nonumber\\
&  =\frac{1}{L^{3}}\sum_{k_{1},k_{2},k_{3}}\frac{1}{e^{2m_{N}\alpha_{t}%
}e^{-ip_{\ast0}}-1+2\omega_{k}-\omega_{k}^{2}}.
\end{align}
Since we are interested in imaginary $p_{\ast0}$ we switch variables,%

\begin{equation}
E+2m_{N}=ip_{\ast0}\alpha_{t}^{-1},
\end{equation}%
\begin{equation}
e^{-ip_{\ast0}}=e^{-\alpha_{t}(E+2m_{N})}\text{,}%
\end{equation}
where $E$ is the energy in Minkowski space, with rest energy excluded.
\ Finally we get%
\begin{equation}
B(E)=\frac{1}{L^{3}}\sum_{k_{1},k_{2},k_{3}}\frac{1}{e^{-\alpha_{t}%
E}-1+2\omega_{k}-\omega_{k}^{2}}, \label{bubble}%
\end{equation}
and the pole in the bubble chain sum occurs when
\begin{equation}
B(E)=\frac{1}{(1-6h)^{2}\left(  e^{-C\alpha_{t}}-1\right)  }\text{.}%
\end{equation}

Using (\ref{lus}) we can determine $C$ for various spatial and temporal
lattice spacings. \ Results for the lattice spacings used in the simulations
presented here are shown in Table 2. \ We have also determined $\frac
{dC}{d\alpha_{t}}$, which will be needed when computing the average energy by
varying $\beta$ with fixed $L_{t}$.%
\[%
\genfrac{}{}{0pt}{0}{\text{Table 2: Contact potential coefficients (MeV}%
^{-2}\text{)}}{%
\begin{tabular}
[c]{|l|l|l|l|}\hline
$a^{-1}($MeV$)$ & $a_{t}^{-1}($MeV$)$ & $C($MeV$^{-2})$ & $\frac{dC}%
{d\alpha_{t}}$\\\hline
$50$ & $24$ & $-11.6\times10^{-5}$ & $-2.21\times10^{-5}$\\\hline
$60$ & $32$ & $-10.1\times10^{-5}$ & $-2.31\times10^{-5}$\\\hline
$60$ & $48$ & $-8.75\times10^{-5}$ & $-1.91\times10^{-5}$\\\hline
$70$ & $64$ & $-7.58\times10^{-5}$ & $-1.92\times10^{-5}$\\\hline
$80$ & $72$ & $-6.96\times10^{-5}$ & $-2.04\times10^{-5}$\\\hline
\end{tabular}
}%
\]

\section{Bubble chain summation}

\begin{figure}
[ptb]
\begin{center}
\includegraphics[
height=1.0888in,
width=2.1136in
]%
{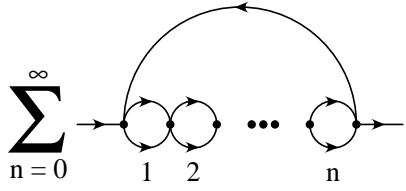}%
\caption{Bubble chain diagrams contributing to the neutron self-energy.}%
\label{self}%
\end{center}
\end{figure}

\begin{figure}
[ptb]
\begin{center}
\includegraphics[
height=1.3206in,
width=1.8922in
]%
{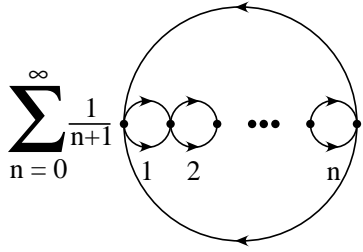}%
\caption{Bubble chain diagrams contributing to the logarithm of the partition
function.}%
\label{energy}%
\end{center}
\end{figure}

In this section we discuss a simple semi-analytic calculation that we use to
compare with the results of our simulations. \ At $T=0$ and if $k_{F}|a_{nn}|$
small the energy and particle densities can be calculated as an expansion in
$k_{F}|a_{nn}|$. \ If the scattering length $a_{nn}$ is small this is
equivalent to a perturbative expansion in the coupling constant $C$. \ If
$a_{nn}$ is not small then an infinite set of particle-particle bubbles has to
be summed. \ This is particularly obvious in the lattice cutoff scheme
employed in this work. \ Since the coupling constant $C$ is fixed by matching
the particle-particle bubble sum to the experimental scattering length at a
given lattice spacing, a perturbative expansion of the equation of state in
powers of $C$ will not be cutoff independent. \ An approximation scheme that
will reproduce the lowest order $k_{F}a_{nn}$ expansion is the bubble chain
summation shown in Figs. \ref{self} and \ref{energy}.

The problem at $T=0$ is that the scattering length is very large, and the
expansion in $k_{F}|a_{nn}|$ is not useful unless the density is extremely
small, $\rho<10^{-4}\rho_{N}$. \ When $k_{F}|a_{nn}|$ is not small then
corrections must be summed to all orders, and it is not obvious that there is
any subset of diagrams that can approximate the full non-perturbative result.
\ We note, however, that the bubble chain diagrams contain as a subset the
diagrams with the minimum number of hole lines. \ These diagrams are summed by
the low density hole line expansion. \ 

The situation is simpler if the temperature $T$ is large compared to the
degeneracy temperature $T_{F}=(3\pi^{2}\rho)^{2/3}/(2m)$. \ In this case a new
length scale, the thermal wavelength or localization length appears
\begin{equation}
\lambda_{T}\sim\sqrt{\frac{1}{2m_{N}T}}\text{.}%
\end{equation}
This length scale acts as an infrared regulator, cutting off long-distance
correlations beyond this scale. \ In particular, it regulates the
neutron-neutron scattering amplitude near threshold by giving the function
$B(E)$ in (\ref{bubble}) a correction of order $O(\lambda_{T}^{-1})$. \ The
net effect is that neutrons now have an effective scattering length of%
\begin{equation}
\left\vert a_{eff}\right\vert \sim\min(|a_{nn}|,\lambda_{T})\text{.}%
\end{equation}
The expansion in $a_{eff}^{3}\rho$ converges as long as $a_{eff}^{3}\rho<1$
which is equivalent to $T>T_{F}$. In the following we compute the bubble chain
diagrams shown in Figs. \ref{self} and \ref{energy}.\ 

%

The bubble chain diagrams in the neutron self-energy form a geometric series.
\ The sum is given by
\begin{equation}
\Sigma(\vec{q})=-(1-6h)^{2}\left(  e^{-C\alpha_{t}}-1\right)  \sum_{\vec{p}%
}\frac{D^{free}(\vec{p}-\vec{q})}{1-(1-6h)^{2}\left(  e^{-C\alpha_{t}%
}-1\right)  B(\vec{p},\mu)}%
\end{equation}
where%
\begin{equation}
B(\vec{p},\mu)=\frac{1}{L^{3}L_{t}}\sum_{\vec{k}}\frac{1}{e^{(m_{N}-\mu
)\alpha_{t}}e^{-ip_{\ast0}/2}e^{-ik_{\ast0}}-1+\omega_{p/2+k}}\frac
{1}{e^{(m_{N}-\mu)\alpha_{t}}e^{-ip_{\ast0}/2}e^{ik_{\ast0}}-1+\omega
_{-p/2+k}}.
\end{equation}
We use this to compute the full neutron propagator%
\begin{equation}
D^{full}(\vec{q})=\frac{D^{free}(\vec{q})}{1-\Sigma(\vec{q})D^{free}(\vec{q}%
)},
\end{equation}
and the average number of neutrons is%
\begin{equation}
A=\frac{1}{\beta}\frac{\partial}{\partial\mu}\ln Z_{G}=2L^{3}\left[
1-\frac{e^{(m_{N}-\mu)\alpha_{t}}}{L_{t}L^{3}}\sum_{\vec{k}}D^{full}(\vec
{k})e^{-ik_{\ast0}}\right]  .
\end{equation}

In a similar fashion we compute the contribution of bubble chain diagrams to
the logarithm of the partition function. \ The relevant diagrams are shown in
Fig. \ref{energy}.%
The factor of $\frac{1}{n+1}$ is the due to the cyclical symmetry of the
diagram. \ We find%
\begin{align}
\ln Z_{G}  &  =\ln Z_{G}^{free}\nonumber\\
&  +\frac{1}{L_{t}L^{3}}\sum_{\vec{p},\vec{q}}\frac{-\ln\left[  1-(1-6h)^{2}%
\left(  e^{-C\alpha_{t}}-1\right)  B(\vec{p}+\vec{q},\mu)\right]
D^{free}(\vec{p})D^{free}(\vec{q})}{B(\vec{p}+\vec{q},\mu)}.
\end{align}
From this we can compute the average energy $E$%
\begin{equation}
E=-\frac{\partial\ln Z_{G}}{\partial\beta}+(-m_{N}+\mu)A,
\end{equation}
where we have subtracted out the rest energy. \ The derivative with respect to
$\beta$ is calculated at fixed $L_{t}$ by varying $\alpha_{t},$%
\begin{equation}
E=-\frac{1}{L_{t}}\frac{\partial\ln Z_{G}}{\partial\alpha_{t}}+(-m_{N}+\mu)A.
\label{Eformula}%
\end{equation}
We must take into account the dependence of $C$ on $\alpha_{t}$, and
$\frac{dC}{d\alpha_{t}}$ for various lattice spacings are shown in Table 2.

\section{Computational methods}

We use the hybrid Monte Carlo (HMC) algorithm \cite{Duane:1987de} to generate
field configurations. \ Roughly 10$^{5}$ five-step HMC trajectories were run,
split across 9 processors running completely independent trajectories.
\ Averages and errors were computed by comparing the results of each
processor. \ While the HMC algorithm has become standard in lattice QCD, it
may not be so well known in the general nuclear theory community. \ We
therefore include a brief overview of the method as applied to our simulation.

We want to sample the partition function
\begin{equation}
Z_{G}\propto\int DsD\psi D\psi^{\ast}\exp\left[  -\psi_{i}^{\ast}%
Q_{ij}^{\prime}(s)\psi_{j}-V(s)\right]
\end{equation}
where the $s_{\alpha}$'s are bosonic fields and the $\psi_{i}$'s are fermionic
fields. \ We use a prime since we will redefine $Q_{ij}^{\prime}$ shortly.
\ We can rewrite this as a bosonic path integral%

\begin{equation}
Z_{G}\propto\int DsD\phi D\phi^{\ast}\exp\left[  -S^{\prime}(\phi,s)\right]
\end{equation}
where%
\begin{equation}
S^{\prime}(\phi,s)=\phi_{i}^{\ast}Q_{ij}^{\prime-1}(s)\phi_{j}+V(s).
\end{equation}
The $\phi_{i}$'s are bosonic fields and are called pseudofermion fields. \ The
partition function can be written as%

\begin{equation}
Z_{G}\propto\int DsDpD\phi D\phi^{\ast}\exp\left[  -H(\phi,s,p)\right]  ,
\end{equation}
where%

\begin{equation}
H(\phi,s,p)=S^{\prime}(\phi,s)+\frac{1}{2}p_{\alpha}p_{\alpha}.
\end{equation}
We note that%
\begin{align}
\frac{\partial S^{\prime}(\phi,s)}{\partial s_{\alpha}}  &  =\phi_{i}^{\ast
}\frac{\partial Q_{ij}^{\prime-1}(s)}{\partial s_{\alpha}}\phi_{j}%
+\frac{\partial V(s)}{\partial s_{\alpha}}\nonumber\\
&  =-\phi_{i}^{\ast}Q_{ij}^{\prime-1}(s)\frac{\partial Q_{jk}^{\prime}%
(s)}{\partial s_{\alpha}}Q_{kl}^{\prime-1}(s)\phi_{l}+\frac{\partial
V(s)}{\partial s_{\alpha}}.
\end{align}

In our case the determinant of $Q^{\prime}$ is real and non-negative. \ We can
therefore replace $Q_{ij}^{\prime}$ by a positive semi-definite Hermitian
matrix $Q_{ij}$, with the same determinant. \ In our case $Q_{ij}^{\prime}$
has the block diagonal structure%
\begin{equation}
Q^{\prime}=%
\begin{bmatrix}
K & 0\\
0 & K
\end{bmatrix}
,
\end{equation}
one block for the up spins and one block for the down spins. \ Clearly
$K_{ij}$ is a matrix with half the dimension of $Q_{ij}^{\prime}$. \ If we let%
\begin{equation}
Q=K^{\dagger}K
\end{equation}
then%
\begin{equation}
Q^{-1}=K^{-1}\left(  K^{\dagger}\right)  ^{-1}%
\end{equation}
and%
\begin{equation}
\det Q=\det Q^{\prime}\text{.}%
\end{equation}

So now instead of $S^{\prime}(\phi,s)$ we can use%
\begin{align}
S(\phi,s)  &  =\phi_{i}^{\ast}Q_{ij}^{-1}(s)\phi_{j}+V(s)\nonumber\\
&  =\xi_{j}^{\ast}(s)\xi_{j}(s)+V(s)
\end{align}
where%
\begin{align}
\xi_{i}  &  =\left(  K^{\dagger}\right)  _{ij}^{-1}\phi_{j},\\
\phi_{i}  &  =\left(  K^{\dagger}\right)  _{ij}\xi_{j}.
\end{align}
When we compute the derivative with respect to $s_{\alpha}$, we have%
\begin{equation}
\frac{\partial}{\partial s_{\alpha}}\left[  K^{-1}(K^{\dagger})^{-1}\right]
=-K^{-1}\frac{\partial K}{\partial s_{\alpha}}K^{-1}(K^{\dagger})^{-1}%
-K^{-1}(K^{\dagger})^{-1}\frac{\partial K^{\dagger}}{\partial s_{\alpha}%
}(K^{\dagger})^{-1}.
\end{equation}
Let us define%
\begin{equation}
\eta_{i}=K_{ij}^{-1}\xi_{j}.
\end{equation}
Then%
\begin{equation}
\phi_{i}^{\ast}\frac{\partial}{\partial s_{\alpha}}\left[  K^{-1}(K^{\dagger
})^{-1}\right]  _{ij}\phi_{j}=-\xi_{i}^{\ast}\left[  \frac{\partial
K}{\partial s_{\alpha}}\right]  _{ij}\eta_{j}-\eta_{i}^{\ast}\left[
\frac{\partial K^{\dagger}}{\partial s_{\alpha}}\right]  _{ij}\xi_{j}.
\end{equation}
Therefore%
\begin{equation}
\frac{\partial S(\phi,s)}{\partial s_{\alpha}}=-\xi_{i}^{\ast}\left[
\frac{\partial K}{\partial s_{\alpha}}\right]  _{ij}\eta_{j}-\eta_{i}^{\ast
}\left[  \frac{\partial K^{\dagger}}{\partial s_{\alpha}}\right]  _{ij}\xi
_{j}+\frac{\partial V(s)}{\partial s_{\alpha}}.
\end{equation}
The steps for the HMC algorithm are now as follows.

\begin{itemize}
\item[Step 1:] Select an arbitrary initial real-valued configuration
$s_{\alpha}^{0}$.

\item[Step 2:] Select a complex-valued configuration $\xi_{j}$ according to
the Gaussian random distribution,%
\begin{equation}
P(\xi_{j})\propto\exp\left[  -\left\vert \xi_{j}\right\vert ^{2}\right]  ,
\end{equation}
and let
\begin{align}
\phi_{i}  &  =\left(  K^{\dagger}\right)  _{ij}\xi_{j},\\
\eta_{i}  &  =K_{ij}^{-1}\xi_{j}.
\end{align}

\item[Step 3:] Select real-valued $p_{\alpha}^{0}$ according to the Gaussian
random distribution%
\begin{equation}
P(p_{\alpha}^{0})\propto\exp\left[  -\frac{1}{2}(p_{\alpha}^{0})^{2}\right]  .
\end{equation}

\item[Step 4:] Let%
\begin{equation}
s_{\alpha}(0)=s_{\alpha}^{0},
\end{equation}
and%
\begin{equation}
\tilde{p}_{\alpha}(0)=p_{\alpha}^{0}-\frac{\varepsilon}{2}\left[  -\xi
_{i}^{\ast}\left[  \frac{\partial K}{\partial s_{\alpha}}\right]  _{ij}%
\eta_{j}-\eta_{i}^{\ast}\left[  \frac{\partial K^{\dagger}}{\partial
s_{\alpha}}\right]  _{ij}\xi_{j}+\frac{\partial V(s)}{\partial s_{\alpha}%
}\right]  _{s=s^{0}},
\end{equation}
for some small positive $\varepsilon$.

\item[Step 5:] For $n=0,1,...,N-1$, let%
\begin{align}
s_{\alpha}(n+1)  &  =s_{\alpha}(n)+\varepsilon\tilde{p}_{\alpha}(n),\\
\tilde{p}_{\alpha}(n+1)  &  =\tilde{p}_{\alpha}(n)-\varepsilon\left[  -\xi
_{i}^{\ast}\left[  \frac{\partial K}{\partial s_{\alpha}}\right]  _{ij}%
\eta_{j}-\eta_{i}^{\ast}\left[  \frac{\partial K^{\dagger}}{\partial
s_{\alpha}}\right]  _{ij}\xi_{j}+\frac{\partial V(s)}{\partial s_{\alpha}%
}\right]  _{s=s(n+1)}.
\end{align}

\item[Step 6:] Let%
\begin{equation}
p_{\alpha}(N)=\tilde{p}_{\alpha}(N)+\frac{\varepsilon}{2}\left[  -\xi
_{i}^{\ast}\left[  \frac{\partial K}{\partial s_{\alpha}}\right]  _{ij}%
\eta_{j}-\eta_{i}^{\ast}\left[  \frac{\partial K^{\dagger}}{\partial
s_{\alpha}}\right]  _{ij}\xi_{j}+\frac{\partial V(s)}{\partial s_{\alpha}%
}\right]  _{s=s(N)}.
\end{equation}

\item[Step 7:] Select a random number $r\in$ $[0,1).$ \ If
\begin{equation}
r<\exp\left[  -H(\phi,s(N),p(N))+H(\phi,s^{0},p^{0})\right]
\end{equation}
then let%
\begin{equation}
s^{0}=s(N).
\end{equation}
Otherwise leave $s^{0}$ as is. \ In either case go back to Step 2.
\end{itemize}

The total number of neutrons, $A$, is%
\begin{align}
A  &  =\frac{1}{\beta}\frac{\partial}{\partial\mu}\ln Z_{G}=2L^{3}-\frac
{1}{\beta}\frac{\int DsDc^{\prime}Dc^{\ast}\frac{\partial S}{\partial\mu}%
\exp\left[  -S\right]  }{\int DsDc^{\prime}Dc^{\ast}\exp\left[  -S\right]
}\nonumber\\
&  =2L^{3}\left[  1-\frac{e^{(m_{N}-\mu)\alpha_{t}}\int DsDc^{\prime}Dc^{\ast
}c_{\uparrow}^{\prime}(\vec{n}+\hat{0})c_{\uparrow}^{\ast}(\vec{n})\exp\left[
-S\right]  }{\int DsDc^{\prime}Dc^{\ast}\exp\left[  -S\right]  }\right]  .
\end{align}
for any lattice site $\vec{n}$. \ Dividing by the volume $V=L^{3}$ gives the
density $\rho$ in lattice units. \ We compute the total energy using
(\ref{Eformula}),%
\begin{align}
E  &  =-\frac{1}{L_{t}}\frac{\partial\ln Z_{G}}{\partial\alpha_{t}}%
+(-m_{N}+\mu)A\nonumber\\
&  =\frac{1}{L_{t}}\frac{\int DsDc^{\prime}Dc^{\ast}\frac{\partial S}%
{\partial\alpha_{t}}\exp\left[  -S\right]  }{\int DsDc^{\prime}Dc^{\ast}%
\exp\left[  -S\right]  }+(-m_{N}+\mu)A,
\end{align}
where we take into account the $\alpha_{t}$ dependence of $C$ when computing
$\frac{\partial S}{\partial\alpha_{t}}$. \ We then have%

\begin{equation}
E=2L^{3}\frac{\int DsDc^{\prime}Dc^{\ast}f(s,c^{\prime},c^{\ast})\exp\left[
-S\right]  }{\int DsDc^{\prime}Dc^{\ast}\exp\left[  -S\right]  }+(-m_{N}+\mu)A
\end{equation}
where
\begin{align}
f(s,c^{\prime},c^{\ast})  &  =-(m_{N}-\mu)e^{(m_{N}-\mu)\alpha_{t}}%
c_{\uparrow}^{\prime}(\vec{n}+\hat{0})c_{\uparrow}^{\ast}(\vec{n})\nonumber\\
&  +\frac{\partial}{\partial\alpha_{t}}\left(  e^{\sqrt{-C\alpha_{t}}s(\vec
{n})+\frac{C\alpha_{t}}{2}}(1-6h)\right)  c_{\uparrow}^{\prime}(\vec
{n})c_{\uparrow}^{\ast}(\vec{n})\nonumber\\
&  +\frac{1}{2m_{N}}\sum_{l_{s}}\left[  c_{\uparrow}^{\prime}(\vec{n}+\hat
{l}_{s})c_{\uparrow}^{\ast}(\vec{n})+c_{\uparrow}^{\prime}(\vec{n}-\hat{l}%
_{s})c_{\uparrow}^{\ast}(\vec{n})\right]
\end{align}
for any lattice site $\vec{n}$.

\section{Free neutron results}

To better understand our lattice discretization errors, we compare our free
neutron results on the lattice with the continuum free Fermi gas. \ For a
continuum free Fermi gas, the logarithm of the partition function is%
\begin{equation}
\ln Z_{G}^{free}=\ln Z_{G,\uparrow}^{free}+\ln Z_{G,\downarrow}^{free}=2\ln
Z_{G,\uparrow}^{free},
\end{equation}
where the logarithm of the single spin partition function is%

\begin{align}
\ln Z_{\uparrow}^{free}  &  =V\int\frac{d^{3}\vec{p}}{(2\pi)^{3}}\ln\left[
1+e^{-\beta\left(  \frac{\vec{p}^{2}}{2m_{N}}+m_{N}-\mu\right)  }\right]
\nonumber\\
&  =\frac{V}{2\pi^{2}}\int_{0}^{\infty}dp\,p^{2}\ln\left[  1+e^{-\beta\left(
\frac{p^{2}}{2m_{N}}+m_{N}-\mu\right)  }\right]  .
\end{align}
Therefore the energy density is%
\begin{equation}
\frac{E_{\uparrow}^{free}}{V}=\frac{1}{2\pi^{2}}\int_{0}^{\infty}dp\frac
{p^{4}}{2m_{N}}\frac{1}{e^{\beta\left(  \frac{p^{2}}{2m_{N}}+m_{N}-\mu\right)
} +1},
\end{equation}
and the number density is%
\begin{equation}
\rho_{\uparrow}^{free}=\frac{A^{free}}{V}=\frac{1}{2\pi^{2}}\int_{0}^{\infty
}dp\,p^{2}\frac{1}{e^{\beta\left(  \frac{p^{2}}{2m_{N}}+m_{N}-\mu\right)  }%
+1}.
\end{equation}
We double these to get the results for both spins. \ In the limit as
$\rho^{free}\rightarrow0$ we find the usual equipartition result for the
energy per neutron,%
\begin{equation}
\frac{E^{free}}{A^{free}}=\frac{3}{2}T.
\end{equation}

\begin{figure}
[ptb]
\begin{center}
\includegraphics[
height=4.4521in,
width=3.1263in,
angle=-90
]%
{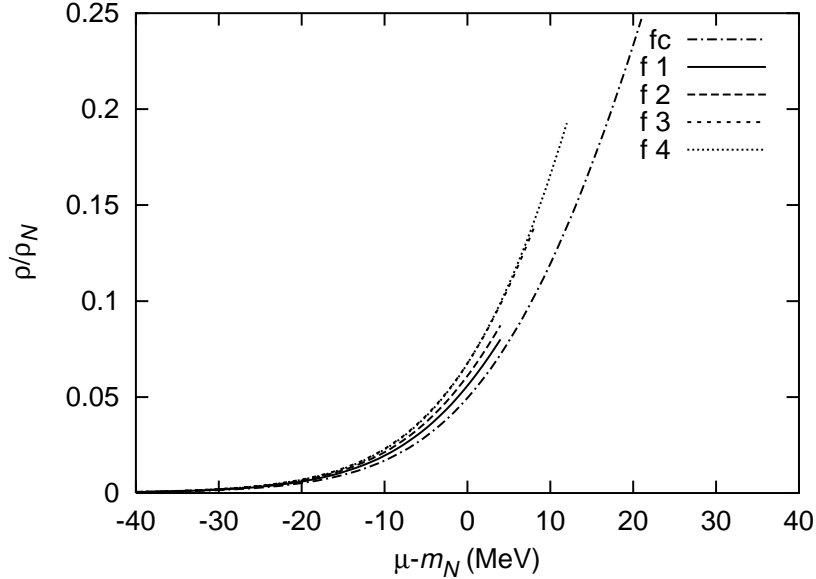}%
\caption{Density versus chemical potential for free neutrons on the lattice at
$T=8$ MeV and various lattice spacings.The curves labeled f1-f4 refer to the
lattice spacings defined in Table 1. The curve labeled fc shows the continuum
limit for free neutrons.}%
\label{Rho_chem_8_raw}%
\end{center}
\end{figure}

\begin{figure}
[ptb]
\begin{center}
\includegraphics[
height=4.4521in,
width=3.1263in,
angle=-90
]%
{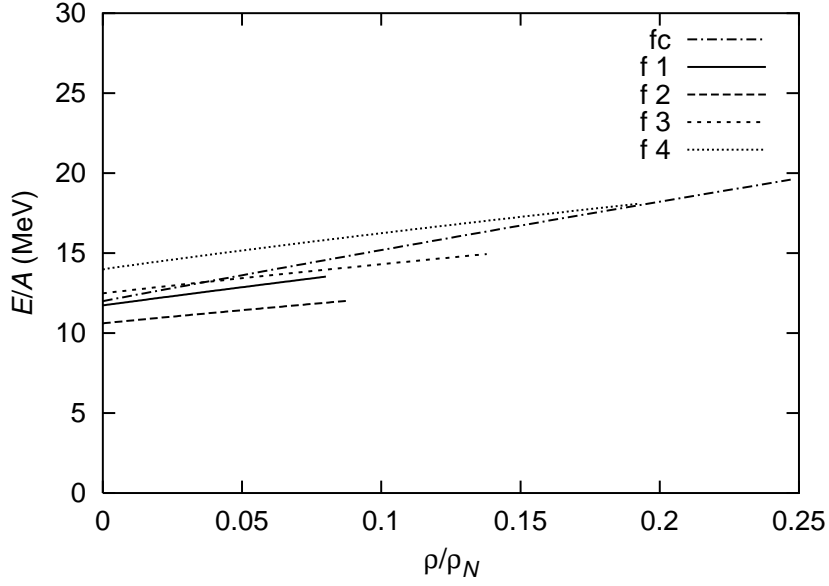}%
\caption{Energy per neutron versus density for free neutrons on the lattice at
$T=8$ MeV and various lattice spacings.}%
\label{Energy_8_raw}%
\end{center}
\end{figure}

A plot of density versus chemical potential at temperature $T=8$ MeV is shown
in Fig. \ref{Rho_chem_8_raw}. \ The energy per neutron at temperature $T=8$
MeV is shown in Fig. \ref{Energy_8_raw}. \ In order to avoid large cutoff
effects, we only present results at densities corresponding with lattice
fillings of about one-quarter or less. \ This is why our data at longer
lattice spacings terminates at lower densities.%
%

We see in Figs. \ref{Rho_chem_8_raw} and \ref{Energy_8_raw} some residual
dependence on lattice spacings. \ While it is a small effect, it does make it
visually confusing to overlay plots for different lattice spacings. \ We
mentioned the possibility of using improved actions to reduce residual lattice
discretization error. \ In this analysis, however, we use a less expensive
route. \ We will simply rescale our densities and energies so that the free
lattice results and free continuum results agree at $\mu=m_{N}$,%
\begin{equation}
\rho(a,a_{t},L,T,\mu)\rightarrow\rho(a,a_{t},L,T,\mu)\cdot\frac{\rho
^{free}(a=0,a_{t}=0,L=\infty,T,\mu=m_{N})}{\rho^{free}(a,a_{t},L,T,\mu=m_{N}%
)}, \label{rho-rescale}%
\end{equation}%
\begin{equation}
E(a,a_{t},L,T,\mu)\rightarrow E(a,a_{t},L,T,\mu)\cdot\frac{E^{free}%
(a=0,a_{t}=0,L=\infty,T,\mu=m_{N})}{E^{free}(a,a_{t},L,T,\mu=m_{N})}.
\label{E-rescale}%
\end{equation}
We will apply the same multiplicative adjustment to all lattice results.
\ This includes free lattice results, bubble chain diagram results, and
lattice simulation results.

\section{Volume dependence}

For the $T=8$ MeV simulations we use a lattice volume of $(13$ fm$)^{3}$ or
larger. \ For the $T=4$ MeV simulations we use a lattice volume of $(20$
fm$)^{3}$ or larger. \ The dimensions of our $L^{3}\times L_{t}$ lattices are
shown in Tables 3 and 4.%

\[%
\genfrac{}{}{0pt}{0}{\text{Table 3: Lattice dimensions for }T=8\text{ MeV}}{%
\begin{tabular}
[c]{|l|l|l|}\hline
Label & $L$ & $L_{t}$\\\hline
$1$ & $4$ & $4$\\\hline
$2$ & $4$ & $6$\\\hline
$3$ & $5$ & $8$\\\hline
$4$ & $6$ & $9$\\\hline
\end{tabular}
}%
\]%
\[%
\genfrac{}{}{0pt}{0}{\text{Table 4: Lattice dimensions for }T=4\text{ MeV}}{%
\begin{tabular}
[c]{|l|l|l|}\hline
Label & $L$ & $L_{t}$\\\hline
$0$ & $5$ & $6$\\\hline
$1$ & $6$ & $8$\\\hline
\end{tabular}
}%
\]
We have run simulations at both smaller and larger volumes. \ In Table 5 we
show the results for $T=8$ MeV and $\mu-m_{N}=-2$ MeV, $a^{-1}=60$ MeV, and
$a_{t}^{-1}=32$ MeV. \ Since we are not changing the lattice spacings for this
comparison we can compare raw data without rescaling $\rho$ and $E.$%
\[%
\genfrac{}{}{0pt}{0}{\text{Table 5: }L\text{ dependence for }T=8\text{ MeV}}{%
\begin{tabular}
[c]{|l|l|l|l|l|l|l|}\hline
$L$ & $\frac{\rho^{free}}{\rho_{N}}$ & $\frac{E^{free}}{A^{free}}$(MeV) &
$\frac{\rho^{bubble}}{\rho_{N}}$ & $\frac{E^{bubble}}{A^{bubble}}$(MeV) &
$\frac{\rho^{simulation}}{\rho_{N}}$ & $\frac{E^{simulation}}{A^{simulation}}%
$(MeV)\\\hline
$3$ & $0.04588$ & $12.582$ & $0.08176$ & $6.514$ & $0.0885(4)$ &
$6.19(2)$\\\hline
$4$ & $0.04596$ & $12.777$ & $0.08215$ & $6.539$ & $0.0890(2)$ &
$6.21(2)$\\\hline
$5$ & $0.04600$ & $12.757$ & $0.08217$ & $6.531$ & $0.0886(3)$ &
$6.19(2)$\\\hline
$6$ & $0.04600$ & $12.756$ & $0.08217$ & $6.531$ & $0.0891(3)$ &
$6.20(2)$\\\hline
\end{tabular}
}%
\]
In Table 6 we show analogous results for $T=4$ MeV and $\mu=m_{N}$,
$a^{-1}=50$ MeV, and $a_{t}^{-1}=24$ MeV.%
\[%
\genfrac{}{}{0pt}{0}{\text{Table 6: }L\text{ dependence for }T=4\text{ MeV}}{%
\begin{tabular}
[c]{|l|l|l|l|l|l|l|}\hline
$L$ & $\frac{\rho^{free}}{\rho_{N}}$ & $\frac{E^{free}}{A^{free}}$(MeV) &
$\frac{\rho^{bubble}}{\rho_{N}}$ & $\frac{E^{bubble}}{A^{bubble}}$(MeV) &
$\frac{\rho^{simulation}}{\rho_{N}}$ & $\frac{E^{simulation}}{A^{simulation}}%
$(MeV)\\\hline
$4$ & $0.02234$ & $7.349$ & $0.04663$ & $3.474$ & $0.0536(3)$ & $3.33(2)$%
\\\hline
$5$ & $0.02238$ & $7.344$ & $0.04667$ & $3.469$ & $0.0533(2)$ & $3.33(2)$%
\\\hline
$6$ & $0.02238$ & $7.341$ & $0.04666$ & $3.469$ & $0.0530(2)$ & $3.35(2)$%
\\\hline
$7$ & $0.02238$ & $7.341$ & $0.04666$ & $3.469$ & $0.0530(2)$ & $3.35(2)$%
\\\hline
\end{tabular}
}%
\]
These results suggest that finite volume effects for the lattice sizes listed
in Tables 3 and 4 are smaller than our statistical errors. \ If we take the
volume dependence from the free and bubble chain calculations as a guide, then
the finite volume errors are well below the $1\%$ level.

\section{Density versus chemical potential}

\begin{figure}
[ptb]
\begin{center}
\includegraphics[
height=4.4521in,
width=3.1263in,
angle=-90
]%
{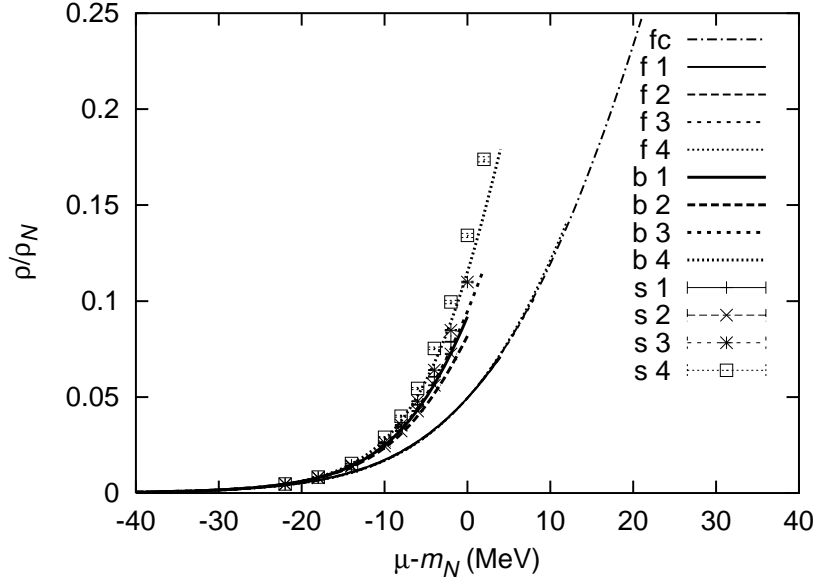}%
\caption{Density versus chemical potential at $T=8$ MeV and various lattice
spacings.The curves labeled f1-f4 show free neutron results, b1-b4 show the
bubble chain results, and s1-s4 show numerical simulations. The corresponding
lattice spacings are given in Table 1.}%
\label{Rho_chem_8}%
\end{center}
\end{figure}

\begin{figure}
[ptbptb]
\begin{center}
\includegraphics[
height=4.0646in,
width=2.8539in,
angle=-90
]%
{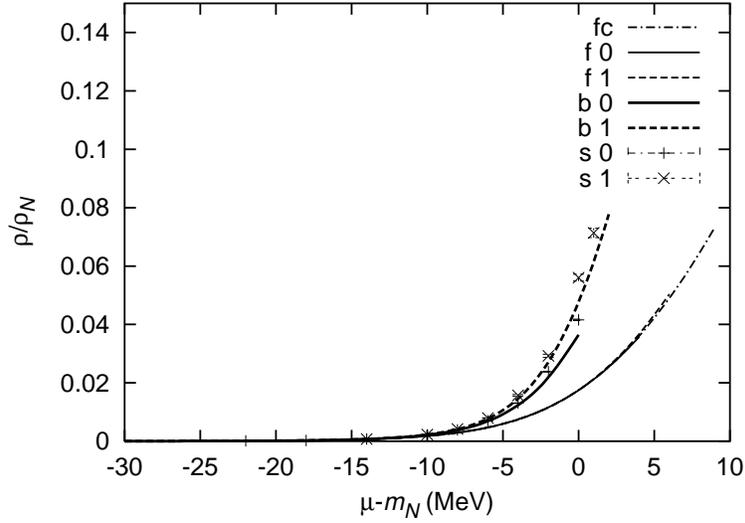}%
\caption{Density versus chemical potential at $T=4$ MeV and various lattice
spacings.}%
\label{Rho_chem_4}%
\end{center}
\end{figure}

In Fig. \ref{Rho_chem_8} we plot density versus chemical potential for $T=8$
MeV, and in Fig. \ref{Rho_chem_4} we plot density versus chemical potential
for $T=4$ MeV.%
In both cases we see agreement among data for different lattice spacings.
\ This suggests that we have properly renormalized the interaction and
absorbed the lattice spacing dependence into the scale dependent interaction
coefficient. \ We observe no phase transitions as a function of chemical
potential. \ We note, in particular, that we can choose the chemical potential
such that the occupation number in the interacting theory remains small.
\ This implies that there is no instability towards a fully occupied ground
state. \ As expected for a theory with attractive interactions the density at
a given chemical potential is larger in the interacting theory. \ We observe
that this behavior is well described by the bubble chain results for $T>T_{F}%
$, the low-density regime where we expect agreement.

\section{Energy per neutron versus density}

\begin{figure}
[ptb]
\begin{center}
\includegraphics[
height=4.0646in,
width=2.8539in,
angle=-90
]%
{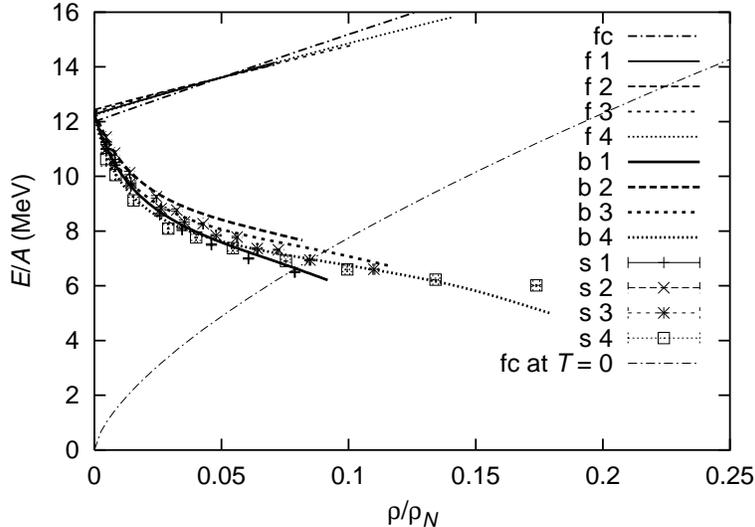}%
\caption{Energy per neutron versus density at $T=8$ MeV for various lattice
spacings.}%
\label{Energy_rho_8}%
\end{center}
\end{figure}

\begin{figure}
[ptbptb]
\begin{center}
\includegraphics[
height=4.0646in,
width=2.8539in,
angle=-90
]%
{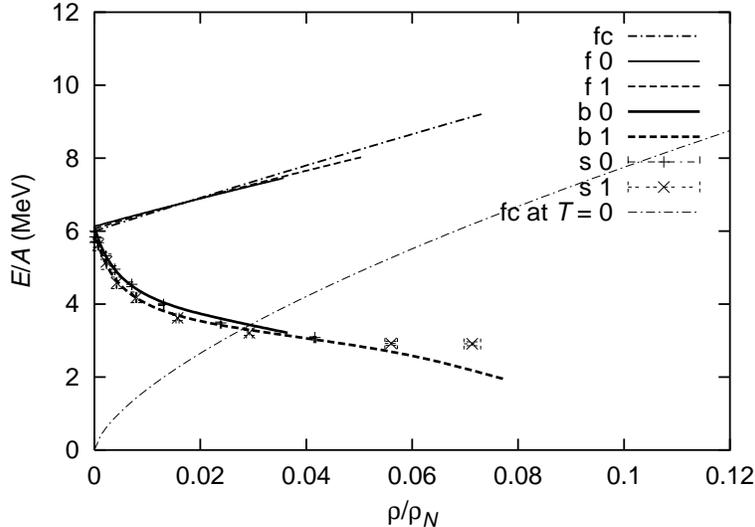}%
\caption{Energy per neutron versus density at $T=4$ MeV for various lattice
spacings.}%
\label{Energy_rho_4}%
\end{center}
\end{figure}

In Fig. \ref{Energy_rho_8} we plot energy per neutron versus density for $T=8$
MeV, and in Fig. \ref{Energy_rho_4} we plot energy per neutron versus density
for $T=4$ MeV.%
In both cases we see good agreement among data for different lattice spacings.
\ As expected, the energy per particle approaches $1.5T$ in the dilute limit.
\ The energy per particle decreases as a function of density in the regime
that we have studied. \ At small $\rho$ the slope is very steep, which is
consistent with the perturbative result for particles that have a large
negative scattering length. \ We note, however, that the behavior is not
linear, even at very small density. \ The simulations are very well described
by the bubble chain results for $T>T_{F}$.\ \ We also observe that for the
larger densities, $\rho>0.1\rho_{N}$ at $T=8$ MeV and $\rho>0.03\rho_{N}$ at
$T=4$ MeV, the energy per particle is smaller than the result for a free
neutron gas at zero temperature. \ In this regime the Fermi energy of the
degenerate system is lower than the temperature.\ Our results suggest that the
parameter $\xi$ defined in (\ref{Bertsch}) is smaller than 0.5. \ We should
note, however, that the parameter $k_{F}r_{nn}\sim k_{F}a$, where $a$ is the
lattice spacing, is of order 1 and simulations at lower temperature will be
required in order to make more definitive estimates of $\xi$.

The decrease in the energy per neutron with increasing density does not
necessarily imply an instability to neutron clustering. \ At nonzero
temperature entropy must also be taken into account, and the question of
whether or not phase separation occurs will be resolved in the next section
when we look at the equation of state.

\section{Equation of state}

\begin{figure}
[ptb]
\begin{center}
\includegraphics[
height=4.4521in,
width=3.1263in,
angle=-90
]%
{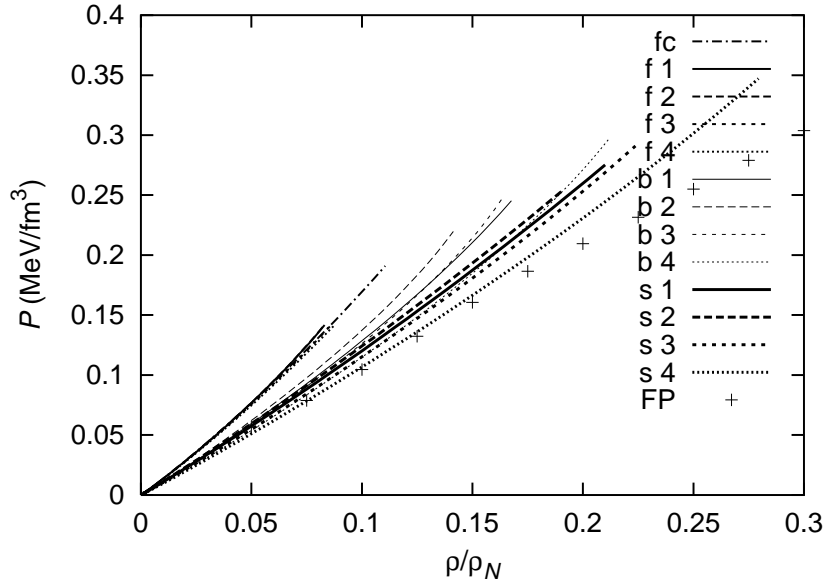}%
\caption{Pressure versus density at $T=8$\ MeV for various lattice spacings.
The crosses show the results of a variational calculation by Friedman and
Pandharipande \cite{Friedman:1981qw}.}%
\label{Pressure8}%
\end{center}
\end{figure}

\begin{figure}
[ptbptb]
\begin{center}
\includegraphics[
height=4.4521in,
width=3.1263in,
angle=-90
]%
{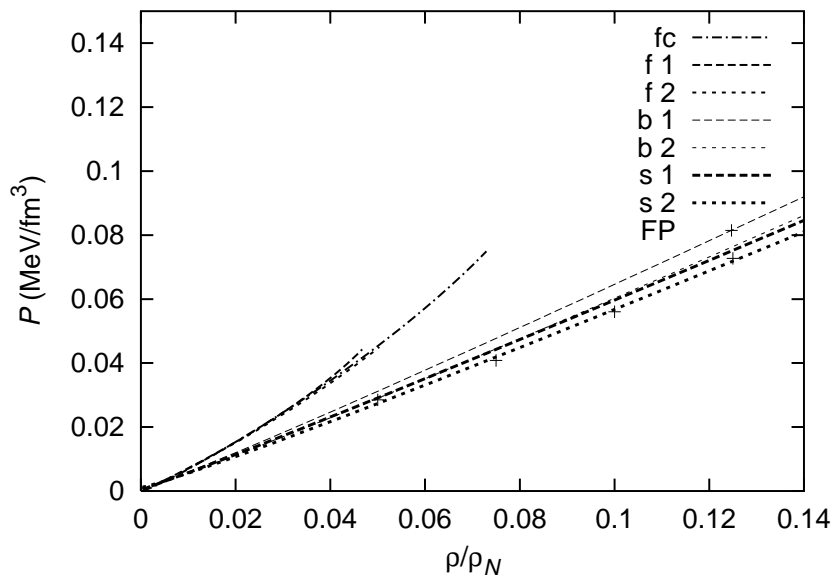}%
\caption{Pressure versus density at $T=4$ MeV for various lattice spacings.
The crosses show the results of a variational calculation by Friedman and
Pandharipande \cite{Friedman:1981qw}.}%
\label{Pressure4}%
\end{center}
\end{figure}

We integrate the density as a function of chemical potential to measure the
pressure,%
\begin{equation}
P=\frac{T}{V}\ln Z_{G}=\frac{1}{V}\int_{-\infty}^{\mu}A(\mu^{\prime}%
)d\mu^{\prime}=\int_{-\infty}^{\mu}\rho(\mu^{\prime})d\mu^{\prime}.
\end{equation}
We perform the integration by least-squares fitting $\rho(\mu^{\prime})$ with
a function of the form
\begin{equation}
\rho(\mu^{\prime})=(c_{0}+c_{1}\mu^{\prime}+c_{2}\mu^{\prime2})\exp
(b\mu^{\prime}).
\end{equation}
We can then perform the integration analytically. \ In Fig. \ref{Pressure8} we
plot the pressure versus density for $T=8$ MeV, and in Fig. \ref{Pressure4} we
plot the pressure versus density for $T=4$ MeV.%
In both cases the pressure is a smooth strictly increasing function of
density. \ Therefore we conclude that there is no indication of phase separation.

There are many models we could use to compare with our results. \ We first
consider the results of a variational calculation by Friedman and
Pandharipande \cite{Friedman:1981qw}. \ They use a realistic Hamiltonian that
consists of the Argonne $v_{14}$ interaction supplemented by a three-body
force. \ We have taken the data for different temperatures given in Table 4 of
\cite{Friedman:1981qw} and interpolated to obtain the pressure for $T=4$ MeV
and $T=8$ MeV. \ The result are shown by the crosses in Figs.~\ref{Pressure8}
and \ref{Pressure4}. \ We observe that the agreement with our calculations is
remarkably good. \ There are a number of factors that are likely to contribute
to this result. \ One is the fact that for the low densities considered in the
present work explicit pions as well as three-body forces are not important.
\ Another point is that we work with relatively coarse lattices. \ On these
lattices the lattice spacing is close to the effective range parameter in
neutron-neutron scattering. \ 

We also show the equation of state for a simple phenomenological model of the
equation of state described in \cite{Li:1997ra} and the review article
\cite{Li:1997px}. \ The model contains a parameterization of the equation of
state of symmetric nuclear matter adjusted to the saturation properties and
the compressibility. \ The authors consider three possible functional forms of
asymmetry,%
\begin{align}
\text{(version 1)\qquad}P_{asy}  &  =2e_{a}\rho_{N}\left(  \frac{\rho}%
{\rho_{N}}\right)  ^{3}\delta^{2},\\
\text{(version 2)\qquad}P_{asy}  &  =e_{a}\rho_{N}\left(  \frac{\rho}{\rho
_{N}}\right)  ^{2}\delta^{2},\\
\text{(version 3)\qquad}P_{asy}  &  =\frac{1}{2}e_{a}\rho_{N}\left(
\frac{\rho}{\rho_{N}}\right)  ^{\frac{3}{2}}\delta^{2},
\end{align}
where $e_{a}\simeq20$ MeV, and the asymmetry parameter $\delta$ is defined as%
\begin{equation}
\delta=\frac{\rho_{n}-\rho_{p}}{\rho_{n}+\rho_{p}}\text{.}%
\end{equation}%
\begin{figure}
[ptb]
\begin{center}
\includegraphics[
height=4.4521in,
width=3.1263in,
angle=-90
]%
{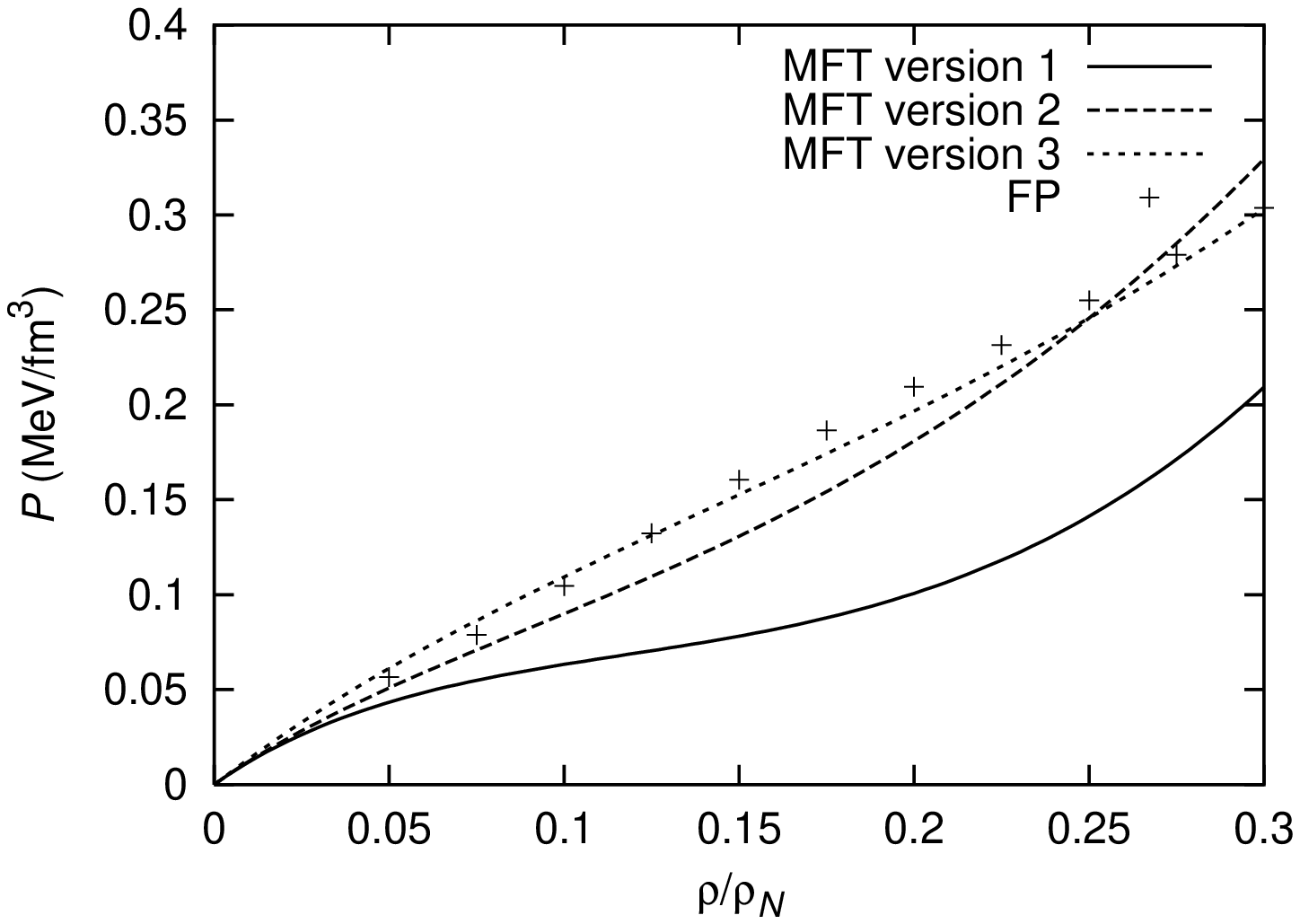}%
\caption{Pressure versus density at $T=8$ MeV for the phenomenological
equation of state discussed in \cite{Li:1997ra}. The three different curves
correspond to different parametrizations of the symmetry energy.}%
\label{MFT8}%
\end{center}
\end{figure}
\begin{figure}
[ptbptb]
\begin{center}
\includegraphics[
height=4.4521in,
width=3.1263in,
angle=-90
]%
{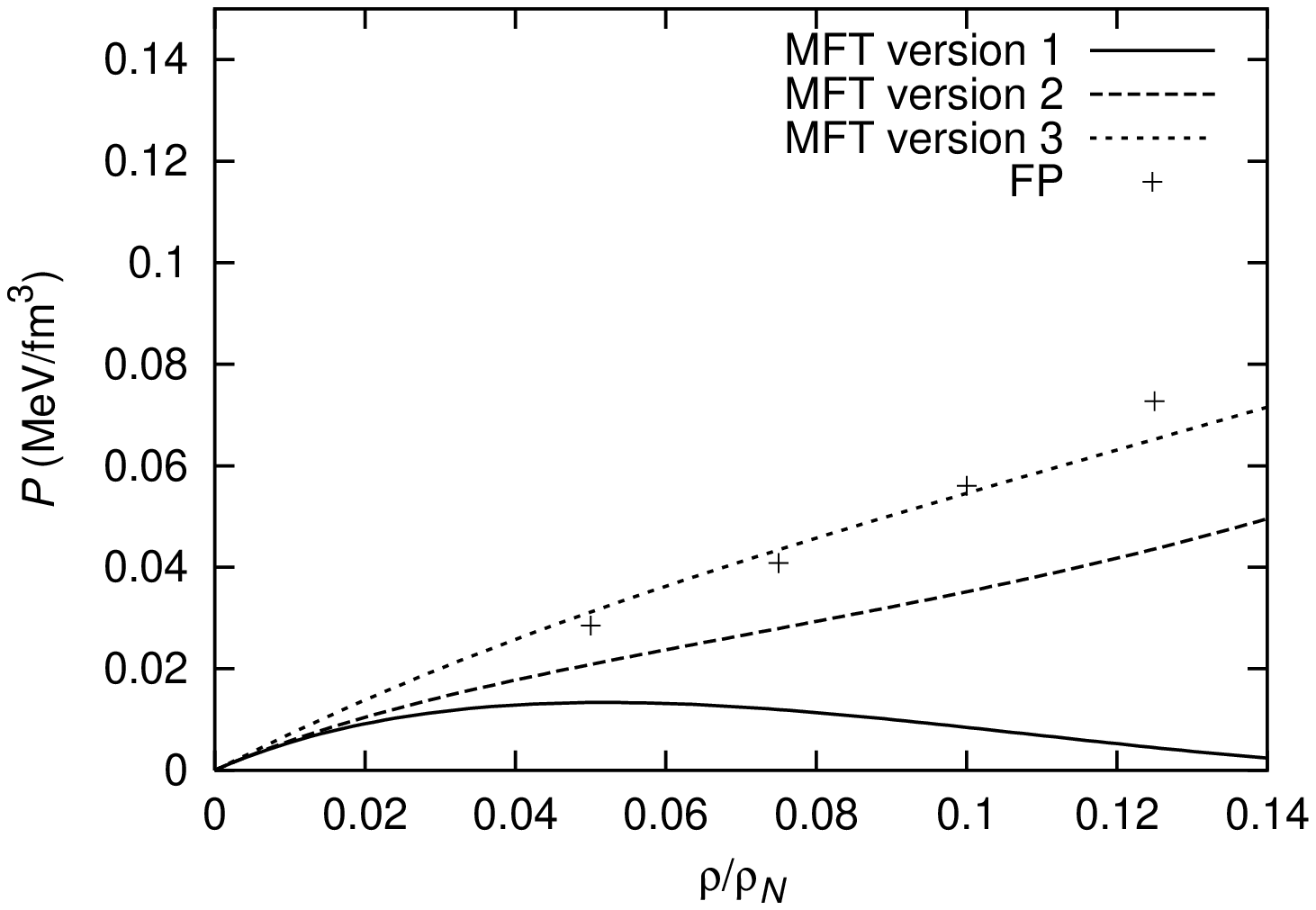}%
\caption{Pressure versus density at $T=4$ MeV for the phenomenological
equation of state discussed in \cite{Li:1997ra}. The three different curves
correspond to different parametrizations of the symmetry energy.}%
\label{MFT4}%
\end{center}
\end{figure}
The equations of state of the three different models for $T=8$ MeV and $T=4$
MeV are shown in Figs. \ref{MFT8} and \ref{MFT4}. \ For comparison, we also
show the variational results of Friedman and Pandharipande. \ We observe that
our results, as well as the results of Friedman and Pandharipande, appear to
agree most closely with version 3. \ Further investigations are needed to
determine whether other properties of this simple model agree with our lattice
simulation results.

\section{Summary and conclusions}

In this work we studied neutron matter by combining pionless effective field
theory at lowest order with non-perturbative lattice methods. \ To determine
the neutron contact interaction we summed bubble chain diagrams contributing
to neutron-neutron scattering at a given lattice spacing. \ The contact
interaction was then adjusted to produce the pole in the amplitude indicated
by L\"{u}scher's finite volume formula for the physical $^{1}S_{0}$ scattering
length.\ Having determined the interaction coefficient for various lattice
spacings, we then simulated neutron matter on the lattice using hybrid Monte
Carlo at temperatures $4$ and $8$ MeV and densities below one-fifth normal
nuclear matter density.

We find that our results at different lattice spacings agree with one another.
\ This suggests that the continuum limit exists and that our effective theory
was properly renormalized or, more conservatively, that any cutoff dependence
is numerically small. \ For the range of parameters studied in this work we
observe no instabilities towards phase separation, or towards lattice
artifacts such as a completely filled lattice. \ While not unexpected, this is
not a trivial result since the non-perturbative simulation includes all
possible diagrams. \ 

The energy per particle at temperatures $T=4$ MeV and $T=8$ MeV shows a steep
downward slope at very small density, which is a sign of the strong attractive
interaction between neutrons.\ At intermediate densities $\rho\sim0.1\rho_{N}$
the energy per particle levels off.\ This behavior is reproduced
quantitatively by our bubble chain calculations for $T>T_{F}$. \ In the future
we wish to push our simulations to lower temperatures and determine the
universal parameters $\xi$ and $\zeta$ for the energy per particle and gap as
defined in (\ref{Bertsch}). \ Simulations in this regime will require a source
term for the di-neutron field. \ We have not seen unambiguous signs of
superfluidity in our simulations. \ We did observe, however, a significant
drop in HMC acceptance rate for the simulation at the lowest temperature and
highest density studied in this work. \ This may well be an indication for the
onset of superfluidity.

Our results for the pressure of pure neutron matter agree remarkably well with
the variational calculation of Friedman and Pandharipande
\cite{Friedman:1981qw}. \ In the future it will be interesting to study
whether the agreement persists if higher order terms in the effective
Lagrangian or explicit pions are introduced. \ We have also performed
simulations with explicit pions \cite{Lee:2004si}, but these data were taken
at larger density and temperature. \ It will also be interesting to study
systems with a finite proton fraction. \ This is easiest in the limit of exact
Wigner symmetry, as the leading order Euclidean action is positive in that
case \cite{Chen:2004rq}. \ 

It will also be interesting to investigate the phase structure of the leading
order effective theory in more detail. \ In order to have a positive Euclidean
action the coefficient $C$ of the four-fermion interaction has to be negative.
\ This corresponds to either a negative scattering length or scattering length
that is large and positive \cite{Chen:2003vy}. \ This implies that the
effective theory studied in this work can be used to investigate the BCS-BEC
crossover in a dilute Fermi gas. \ 

\begin{acknowledgments}
The authors thank Simon Hands and Matthew Wingate for discussions on the
hybrid Monte Carlo algorithm. \ This work was supported in part by DOE grants
DE-FG-88ER40388 and DE-FG02-04ER41335.
\end{acknowledgments}

\bibliographystyle{h-physrev3}
\bibliography{NuclearMatter}

\end{document}